\documentclass[journal=jacsat,manuscript=article]{achemso}
\usepackage[version=3]{mhchem} 

\author{Koushik Karmakar}
\affiliation[Indian Institute of Science Education and Research, Pune, India]
{Indian Institute of Science Education and Research, Pune, Maharashtra-411008, India}
\author{Amartya Singh}
\affiliation[Indian Institute of Science Education and Research, Pune, India]
{Indian Institute of Science Education and Research, Pune, Maharashtra-411008, India}
\author{Surjeet Singh}
\email{surjeet.singh@iiserpune.ac.in}
\phone{+91 20 2590 8049}
\fax{+91 20 2590 8186}
\affiliation[Indian Institute of Science Education and Research, Pune, India]
{Indian Institute of Science Education and Research, Pune, Maharashtra-411008, India}
\author{Amy Poole}
\affiliation[LNS, Paul Scherrer Institute, Switzerland]
{Laboratory for Neutron Scattering, Paul Scherrer Institute, 5232 Villigen PSI, Switzerland}
\author{Christian R\"uegg}
\affiliation[LNS, Paul Scherrer Institute, Switzerland]
{Laboratory for Neutron Scattering, Paul Scherrer Institute, 5232 Villigen PSI, Switzerland}
\alsoaffiliation[DPMC-MaNEP, University of Geneva, Switzerland]
{DPMC-MaNEP, University of Geneva, CH 1121 Geneva, Switzerland}
\title{Crystal growth of the non-magnetic Zn$^{2+}$ and magnetic Co$^{2+}$ doped quasi one-dimensional spin chain compound SrCuO$_{2}$ using the travelling solvent floating zone method}
\begin{document}
\begin{abstract}
We report on the  crystal growth of the quasi one-dimensional quantum spin chain compound SrCuO$_{2}$ and its doped variants containing magnetic cobalt (0.25\%, 0.5\%, 1\% and 2.5\%) and non-magnetic zinc (0.5\% and 1\%) impurities. Crystals are grown using the travelling solvent floating zone (TSFZ) method in a four-mirror optical furnace. Some crucial factors, key to the stability of a TSFZ process, including the choice of solvent composition and its associated melting behavior, are discussed in detail. The grown crystals were characterized using x-ray powder diffraction, scanning electron microscopy, optical microscopy, neutron single crystal diffraction and magnetic susceptibility measurements. Co-doping induces magnetic anisotropy and a corresponding change of magnetic behavior characterized by the presence of a sharp peak in the magnetic susceptibility at low-temperatures (T = 5 K), which is not seen for the pristine compound. The peak position is shown to scale linearly with Co-concentration for low doping levels upto 2.5\%.
\end{abstract}

\section{Introduction}
\begin{figure}[hbtp]
\centering
\includegraphics[width=0.6\textwidth]{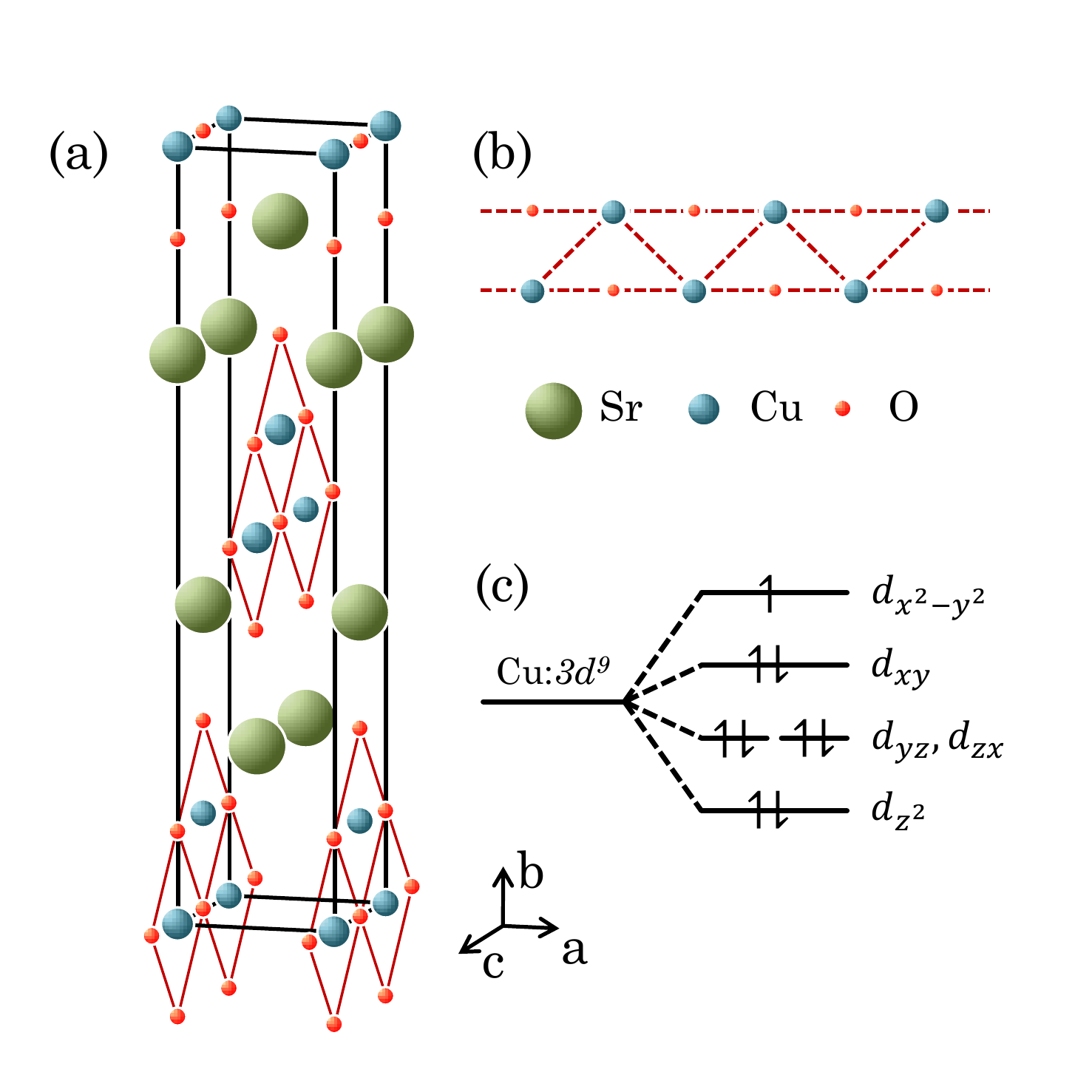}
\caption{(a) Orthorhombic unit cell of SrCuO$_{2}$ (b) Zigzag chain along the c-axis. (c) Crystal field energy level diagram of Cu 3d$^{9}$ in a square-planar geometry.}
\label{crystruc}
\end{figure}

Bulk crystals are inherently three-dimensional, however, they may comprise of magnetic ions whose spins interact only along a certain crystallographic direction. Such compounds are called one-dimensional (1D) magnets or more succinctly 1D spin chains. For small spin values, e.g., S = $\frac{1}{2}$ (as in Cu$^{2+}$ or (low spin) Co$^{2+}$ containing spin chains) or S =1 (as in Ni$^{2+}$ containing spin chains), quantum effects play a prominent role. The name 1D quantum magnet is therefore often used to describe these materials. 1D quantum magnets consisting of antiferromagnetically coupled isotropic S = $\frac{1}{2}$ spins are of considerable interest because of their physical properties, including gapless fractional spin excitations called spinons \cite{Zaliznyak} and the separation of spin, charge and orbital degrees of freedom due to 1D quantum confinement \cite{neudart,kim,kim2,lante,Schalappa}. The thermal conductivity of 1D spin chains has also generated considerable interest. It is characterized by a significant spinon contribution along the chain direction and results in an highly anisotropic behavior (for a review, see for example \cite{hess2}). The model that most aptly captures the physics of 1D quantum magnets is the nearest-neighbor 1D S = $\frac{1}{2}$ Heisenberg antiferromagnetic model (HAM): $H=J\Sigma_{n}(S_{n}.S_{n+1})$ (where n designates a spin site). Quasi-1D SrCuO$_{2}$, which is the subject of the present paper, is considered as one of the best realizations of a S = $\frac{1}{2}$ HAM. It crystallizes with an orthorhombic structure (Fig. \ref{crystruc}a) where each Cu$^{2+}$ ion is in a 4-fold square-planar coordination. In a crystalline electric field of such low symmetry, the orbitally degenerate 3d$^{9}$ level of a free Cu$^{2+}$ ion splits into four sub-levels as shown in fig. \ref{crystruc}b. The orbital $e_{g}(d_{x^{2}- y^{2}})$, which experiences the maximum electrostatic repulsion from the surrounding ligand ions, constitutes the highest energy level with exactly one electron per orbital, which results in a S = $\frac{1}{2}$ chain parallel to the crystallographic c-axis. The absence of intervening oxygen-ions along the a-axis and the conspicuous presence of buffer layers containing magnetically inactive Sr$^{2+}$ perpendicular to the \emph{b}-axis secures magnetic isolation and hence the \emph{magnetic} 1D nature of the compound. As evident from Fig. 1, the spin chains in SrCuO$_{2}$ are not exactly linear but zigzag, which can be thought of as combination of two parallel chains (i.e., a double-chain) that are coupled via a weak 90$^{\circ}$ ferromagnetic superexchange (Fig. 1c). The ferromagnetic exchange is expected to be about an order of magnitude weaker than the dominant 180$^{\circ}$ antiferromagnetic superexchange (J/k$_{B}\sim$ 2000 K)\cite{Rice,Zaliznyak1} . Thus, good one-dimensionality is expected over a broad temperature range. Indeed, the measured spin-excitation spectrum of SrCuO$_{2}$ at T = 12 K shows a gapless two-spinon continuum characteristic of S = $\frac{1}{2}$ HAM spin chains \cite{Zaliznyak}.

Impurities can significantly influence the physical properties of a low-dimensional quantum magnet which has become a major point of discussion recently (for a review, see for example \cite{Alloul}). More specifically, in the context of the compound SrCuO$_{2}$, very recent studies showed that nickel (Ni$^{2+}$: S = 1) impurities ($\sim$1 atomic \%) cause a pseudo-gap in the spin excitation spectrum absent in the parent compound \cite{Simutis}. The thermal conductivity, along the chain direction, is also shown to be extremely sensitive to the presence of impurities in the grown crystal. In particular, signatures of ballistic heat transport are seen in ultra pure crystals of SrCuO$_{2}$ in agreement with the fundamental conservation laws in the integrable 1D S = $\frac{1}{2}$ HAM \cite{Hlubek}. Motivated by these recent developments, we have grown large high-quality single crystals of pure and cobalt (Co$^{2+}$) and zinc (Zn$^{2+}$) doped SrCuO$_{2}$ for inelastic neutron scattering and thermal transport studies.        

The compound SrCuO$_{2}$ exhibits an incongruent melting behaviour; upon heating, it decomposes via a peritectic-type reaction, therefore, conventional growth techniques of obtaining large single-crystals from the corresponding stoichiometric melt cannot be used. The flux growth method using self-flux (CuO) has been shown to work, however, this process is not very useful in view of the size of the grown single crystals (0.2 $\times$ 0.2 $\times$ 2 mm) \cite{Matsushita,Lin}. Moreover, the grown crystal may incorporate traces of the solvent or of the crucible material which would be highly undesirable, particularly, if the aim is to study the effects of specific impurities on the physical properties. With this view in mind we decided to grow crystals of pure and doped SrCuO$_{2}$ using the travelling-solvent-floating-zone (TSFZ) technique associated with an optical image furnace. In this paper, we report the successful growth of large and high-quality single crystals of pristine SrCuO$_{2}$ and those doped with non-magnetic zinc (0.5 and 1 \%) and magnetic cobalt (0.25, 0.5, 1 and 2.5 \%) impurities at the Cu site. We show that by controlling the starting conditions in a TSFZ growth experiment, specially the solvent composition and the initial stability of the floating zone, uninterrupted growth of several cm long crystal at a slow growth rate of about 1 mm/h can be achieved. Our preliminary magnetic study of Co-doped single-crystals along the three principal crystallographic directions have revealed highly anisotropic magnetic response with easy-axis along the chain direction and a prominent peak in the susceptibility at low temperatures (near T = 5 K). The temperature at which the peak appears is shown to scale almost linearly with the Co-concentration.

\section{Experimental Section}
\begin{figure}[hbtp]
\centering
\includegraphics[width=0.6\textwidth]{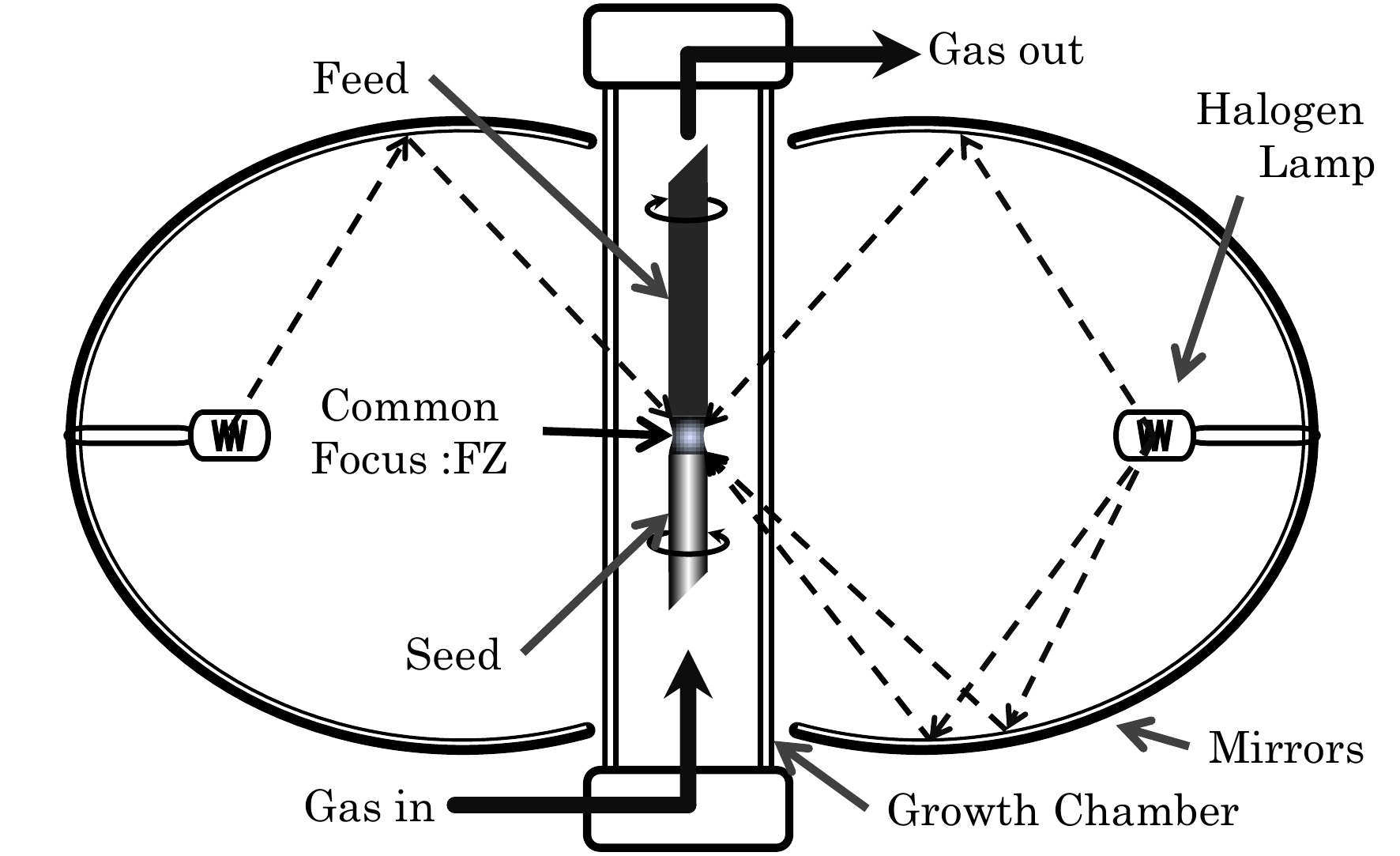}
\caption{General schematic of the vertical section of the optical furnace used in the present work.}
\label{Imagefurnace}
\end{figure}

The single crystals were grown in a four-mirror optical-floating-zone Furnace (from Crystal System Corporation, Japan). A schematic drawing of a vertical section of an optical floating zone furnace is shown in Fig. \ref{Imagefurnace}. The furnace consists of four halogen lamps, each placed at the focus of an ellipsoidal shape reflector. The second focus of these reflectors coincides at a point on the vertical axis of the furnace inside the growth chamber. The energy flux of the lamps, after reflection from the walls of the reflectors, converges at their common foci where a molten zone between the polycrystalline feed and the seed rods is formed. The feed rod is suspended from the upper shaft of the furnace using a nickel wire and the seed rod is clamped to the lower shaft using an alumina holder. For the crystal growth experiments polycrystalline feed rods of compositions: SrCuO$_{2}$ (Pure), SrCu$_{0.995}$Zn$_{0.005}$O$_{2}$ (Zn 0.5\%), SrCu$_{0.99}$Zn$_{0.01}$O$_{2}$ (Zn 1\%), SrCu$_{0.9975}$Co$_{0.0025}$O$_{2}$ (Co 0.25\%), SrCu$_{0.995}$Co$_{0.005}$O$_{2}$ (Co 0.5\%), SrCu$_{0.99}$Co$_{0.01}$O$_{2}$ (Co 1\%) and SrCu$_{0.975}$Co$_{0.025}$O$_{2}$ (Co 2.5\%) were prepared using the standard solid state reaction route. Fine Powders ($<10$ $\mu$m) of SrCO$_{3}$, CuO, ZnO and Co$_{3}$O$_{4}$ were used as the starting precursors. SrCO$_{3}$ was preheated and dried over night at 900$^{\circ}$C before weighing. The powders of the starting precursors were mixed thoroughly in a stoichiometric ratio using a mortar and pestle and then calcined in an alumina crucible at 850$^{\circ}$C. In the subsequent steps, the mixtures were reground and heat treated for 12-24 hours under atmospheric air, with a sequential increase in temperature by 25 to 30$^{\circ}$C at each step with a final heat treatment at 975$^{\circ}$C . The phase analysis of the final product was done using powder x-ray diffraction (Bruker D8 Advance). The powders obtained after the final round of heat treatment were filled in rubber tubes and cold-pressed under 700 bar isostatic pressure to obtain uniform rods measuring about 5 mm in diameter and 80-100 mm in length. The rods were then sintered at temperatures between 985 and 995$^{\circ}$C to obtain very dense, homogeneous rods, which is a necessary condition to ensure uniform melting behaviour during the growth process. About 20 mm long sections, cut from these rods, were used as seeds in the first growth experiments. Table \ref{Synthesis} provides a summary of the synthesis conditions, including the purity of the starting precursors, total number of sintering hours and the results of phase analysis of the final products. 

 The growth experiment in each case is initiated using the corresponding solvent taken in the form of a disk prepared as a pellet of thickness 3 mm. The solvent composition for the undoped compound was taken to be SrO : CuO = 3 : 7 (henceforth, unless mentioned otherwise, the ratio in the context of solvent composition stands for molar ratio SrO : CuO) in accordance with the phase diagram in reference \cite{PD1,PD2}. In the growth of doped compounds, the corresponding amount of dopant was added to the solvent disk so that the doping level in the growing crystal remains equal to that in the feed rod. For example, in the growth of SrCu$_{0.99}$Co$_{0.01}$O$_{2}$, the composition of the solvent used was SrO : CuO : CoO = 3 : 6.93 : 0.07. To test the growth behavior using a different starting solvent composition, we also conducted a growth experiment using a solvent composition 14 : 24 (i.e., a composition coincident with that of the phase Sr$_{14}$Cu$_{24}$O$_{41}$) in accordance with the phase diagram of reference \cite{PD3,PD4}. 
 
The growth process is initiated by first fusing the solvent disk to the bottom end of the feed rod inside the optical furnace. The crystals were grown under O$_{2}$ flow at P = 1 bar. To improve temperature and compositional homogeneity of the float-zone, the upper and lower shafts of the furnace were rotated in opposite directions with rotation speed ranging from 10 to 20 rpm. The growth rate employed in most of the experiments was around 1 mm/h. Often the upper shaft was translated at a speed slower than the lower shaft by 0.1 to 0.2 mm/h in order to control the size of the grown crystal and that of the steady-state floating-zone (Table \ref{GrowthProc}).  

The grown crystals were characterised using powder x-ray diffraction (Bruker D8 Advance), scanning electron microscopy equipped with an energy dispersive X-ray spectrometer (Ziess ultra plus), Laue X-ray diffraction (Photonic Science) and neutron diffraction (MORPHEUS at the Laboratory for Neutron Scattering, PSI, Switzerland). The magnetization of the crystals was measured by applying an external field along the three different crystallographic directions using a vibrating sample magnetometer (VSM, Quantum Design-PPMS-Evercool II) in the temperature range T = 2 to 300 K and applied magnetic field (H) ranging from -80 to +80 kOe. 

\section{Results and Discussion}
\subsection{Crystal growth}
\begin{figure}[hbtp]
\centering
\includegraphics[width=0.6\textwidth]{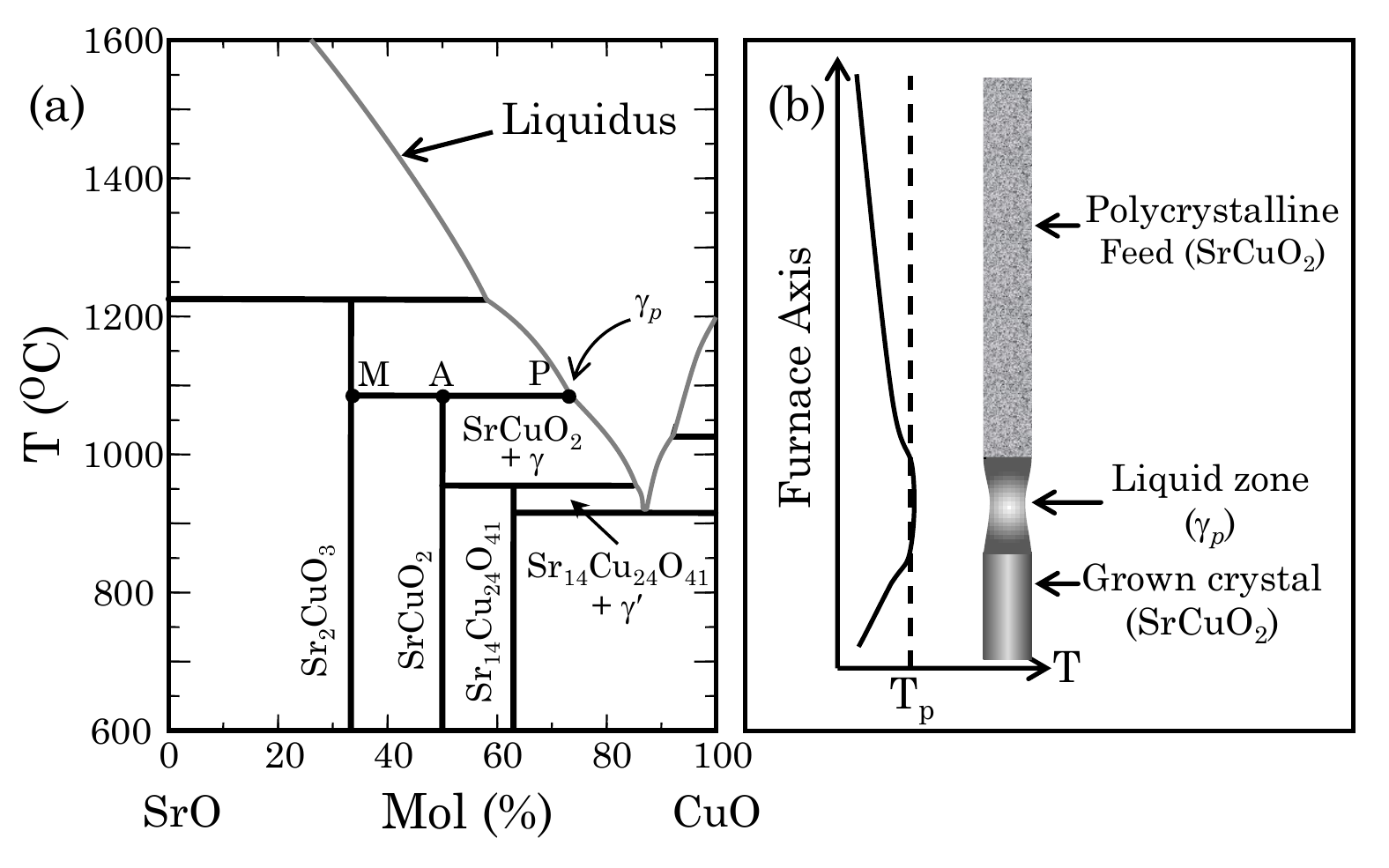}
\caption{Schematic views showing (a) the SrO-CuO phase diagram indicating peritectic decomposition of SrCuO$_{2}$ at temperature T$_{p}$ (adapted from Ref. \cite{PD1}), (b) the TSFZ crystal growth process of SrCuO$_{2}$.}
\label{NCM}
\end{figure}

In the SrO-CuO binary system, three stable incongruently melting compounds having compositions Sr$_{2}$CuO$ _{3}$, SrCuO$_{2}$ and Sr$_{14}$Cu$ _{24} $O$_{41}$ are reported \cite{PD1,PD2,PD3,PD4}.  A schematic drawing of the  SrO-CuO binary phase diagram, showing the peritectic-type decomposition of the compound SrCuO$_{2}$ is presented in Fig. 3a. As shown, upon heating up to a temperature of about T$_{P}$ (1085 +/- 25$^{\circ}$C), the compound decomposes according to the following equation:\\
\begin{equation}
\mathrm{SrCuO}_{2} \xrightarrow{T_{P}} \mathrm{Sr}_{2}\mathrm{CuO}_{3} + \gamma_{p}\mathrm{(L)}
\end{equation} 
where $\gamma_{p}$(L) is the liquid corresponding to the peritectic point (labelled P) in Fig. \ref{NCM} a. At the peritectic temperature, $\gamma_{p}$(L) exists in equilibrium with the solid phase Sr$_{2}$CuO$_{3}$ and their relative molar percentages is given by the Lever rule; i.e., m(S)/m(L) = MA/AP, where m is the number of moles and MA and PA are the intercepts (see Fig. \ref{NCM}a). Because of the peritectic decomposition, the conventional floating zone method [for a review, see for example the article by Koohpayeh et al.\cite{Koohpayeh}] cannot be used here. In such cases, one must resort to the so-called Travelling-Solvent-Floating-Zone (TSFZ) method \cite{Revcolevschi}. As the name suggests, TSFZ is an adaptation of the solvent growth method in an image furnace, where the floating-zone composition and temperature corresponds to that of the associated peritectic point (Fig. \ref{NCM}b). During a continuous TSFZ process, which involves slow downward translation of the upper and the lower shafts, the feed rod decomposes peritectically (eq. 1) at the upper interface of the floating zone and, simultaneously, crystallization of the same number of moles takes place at the lower interface, such that in a steady state condition the volume and the composition of the floating-zone remains constant. An artistic view of the process is depicted in Fig. \ref{NCM}c. In order to attain the steady state condition quickly during a TSFZ process, one usually initiates the growth by melting an appropriate size sintered pellet between the feed and the seed. Ideally, the composition of the pellet should be the same as that of the associated peritectic point \cite{Revcolevschi}.
 
In the particular case of the compound SrCuO$_{2}$, at least four different phase diagrams were previously reported \cite{PD1,PD2,PD3,PD4}. While these phase diagrams are in qualitative agreement with each other; they do not agree about the position of the peritectic point associated with the decomposition of SrCuO$_{2}$. For instance, the phase diagrams due to Liang et.al.\cite{PD3} and Wong et al.\cite{PD4} show the position of the peritectic point at 14 : 24, whereas, Solbodin et al.\cite{PD2} and Hwang et al.\cite{PD1} reported the composition to be 3 : 7. It is worth pointing out here that while the liquidus lines in the phase diagrams due to Wong et al. and Liang et al. appear to be tentative, both, Solbodin et al. and Hwang et al. mark the liquidus and the positions of the peritectic points with more certainty using the data points obtained from phase analysis by X-ray powder diffraction of samples quenched from temperatures as high as 1500$^{\circ}$C.

In previous work dealing with the TSFZ growth of undoped SrCuO$_{2}$ crystals, use of three different solvent compositions has been reported. Ohashi et al.\cite{Ohashi} reported relatively small size crystals, measuring 1 cm in length, using the solvent composition 3 : 7. They also reported floating-zone instabilities for growth speeds exceeding 0.5 mm/h. Revcolevschi et al. \cite{Revcolevschi}, on the other hand, reported several cm long crystals, using growth speed of 1 mm/h and solvent compositions based on the SrO-CuO phase diagrams due to both Hwang et al.\cite{PD1} (3 : 7) and Liang et al. \cite{PD3} (14 : 24). However, they did not publish the details of their growth experiments. 
 Finally, the use of CuO as solvent disk is reported in Ref. \cite{patrickthesis} . The authors pointed out excessive bubbling due to the CuO phase transforming to Cu$_{2}$O + $\frac{1}{2}$O$_{2}$ during melting of the solvent disk; producing an excess of copper in the floating zone. It should be monitored continuously for about 12 hours, during which the temperature-composition variations should closely follow the liquidus until the steady state growth condition is reached. Clearly, the use of a CuO pellet as the starting solvent is not only labour intensive but also hazardous to the quartz enclosure as splintered CuO drops fuse with quartz damaging it permanently. In Ref. \cite{Patrickdip} , where the growth is initiated using a CuO disk, the authors also determined the steady state floating zone composition  using energy dispersive X-ray analysis of the quenched floating zone at the end of a stable growth. Their analysis revealed that under steady state condition the floating zone has a composition of 3 : 7.

\begin{figure}[hbtp]
\centering
\includegraphics[width=0.5\textwidth]{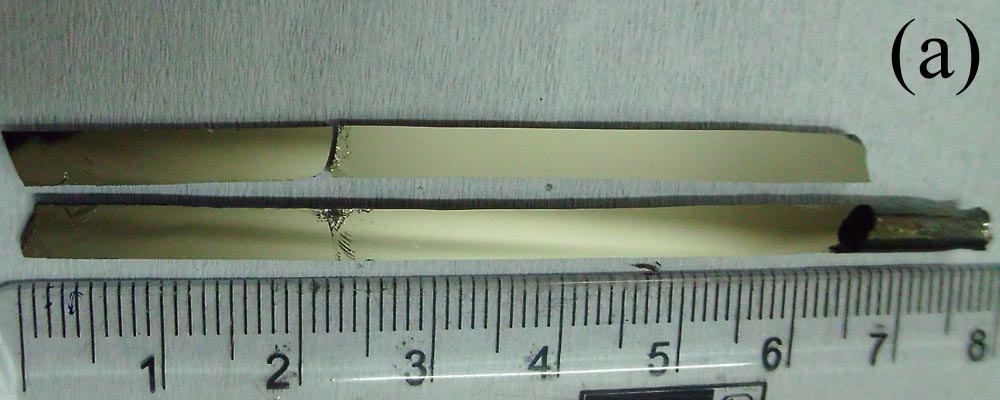}\includegraphics[width=0.5\textwidth]{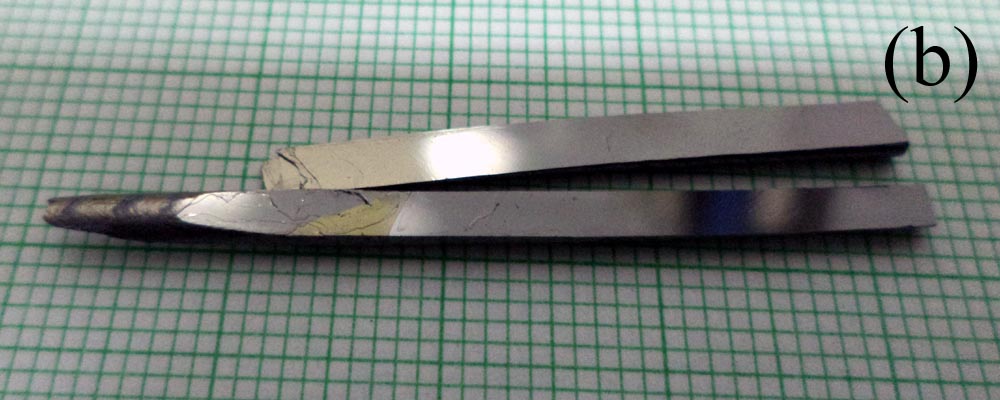}
\includegraphics[width=0.5\textwidth]{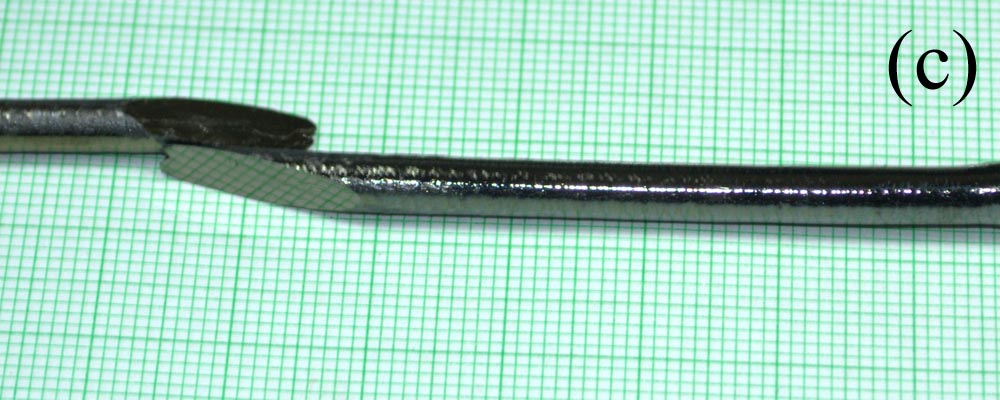}\includegraphics[width=0.5\textwidth]{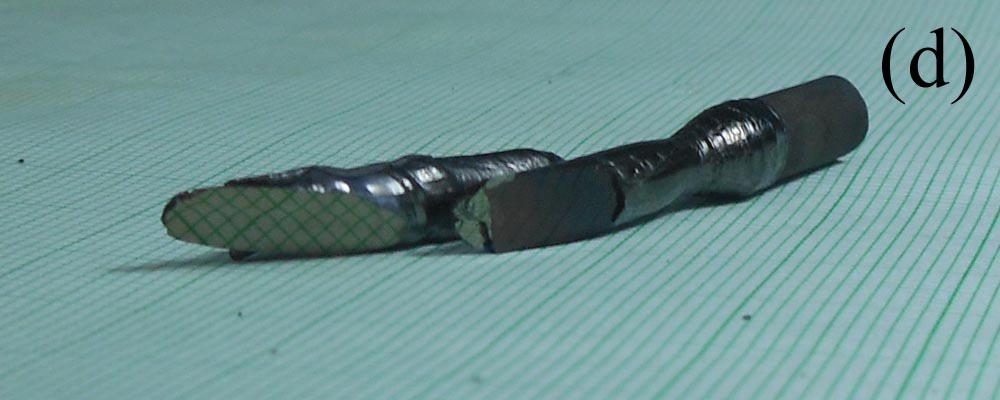}
\caption{Representative images of the as-grown crystals exhibiting mirror finished cleaved surfaces (a) pure (b) Zn1\% (c) Co0.25\% (d) Co2.5\%.}
\label{SingleCry}
\end{figure}

Following the more definitive phase diagrams published in references \cite{PD1,PD2} and the results due to Ref. \cite{Patrickdip} , in the present work we used the solvent composition 3 : 7. Moreover, to confirm that this is indeed the right choice to reach the steady state soon after the floating zone is formed we also tried the composition 14 : 24 in one of the growth experiments. We find that it is possible to reach the steady state condition directly using the solvent composition 3 : 7 but not with the composition 14 : 24. Representative pictures of some of the grown crystals using the solvent composition 3 : 7 are shown in Fig. \ref{SingleCry}. Values of various growth parameters involved in a TSFZ process are summarized in Table \ref{GrowthProc}. 

Having the right solvent composition, therefore, aids in reaching the steady state condition quickly. However, there is another important consideration, whose non-compliance may result in the interruption of the growth process or may introduce large delays in reaching the steady state in spite of starting with the right solvent composition. This concerns the melting behaviour of the solvent pellet. A sintered solvent pellet of compositions 3 : 7 essentially consists of a mixture of solid phases Sr$_{14}$Cu$_{24}$O$_{41}$ and CuO (see Fig. 3b), i.e.,

Sintered solvent pellet $\equiv$ Sr$_{14}$Cu$_{24}$O$_{41}$(S)+CuO(S).

Upon heating, the pellet will undergo successive phase transformations according to the following equation:
\begin{equation}
\mathrm{Sr}_{14}\mathrm{Cu}_{24}\mathrm{O}_{41}\mathrm{(S)} + \mathrm{CuO(S)} \xrightarrow{980^{\circ}\mathrm{C}} \mathrm{Sr}_{14}\mathrm{Cu}_{24}\mathrm{O}_{41}\mathrm{(S)} + \gamma'\mathrm{(L)} \xrightarrow{1003^{\circ}\mathrm{C}} \mathrm{SrCuO}_{2} + \gamma\mathrm{(L)} \xrightarrow{1085^{\circ}\mathrm{C}} \gamma_{p}\mathrm{(L)}
\end{equation}

where $\gamma'$(L) and $\gamma$ (L) are the liquids in equilibrium with solids Sr$_{14}$Cu$_{24}$O$_{41}$ and SrCuO$_{2}$ (Fig. 3a).

From the sequence of transformations that the solvent disk exhibits upon heating, it is evident that unless melted completely the resulting floating zone will neither have the desired composition nor the volume for a sustainable TSFZ process. Very often in the actual experiments, once the decomposition of the solvent disk starts the liquid thus produced tends to flow downward under gravity, which prompts the crystal grower to establish a floating zone by lowering the feed and allowing it to come in contact with a liquid which is more copper rich than desired.

\begin{figure}[hbtp]
\centering
\includegraphics[width=0.7\textwidth]{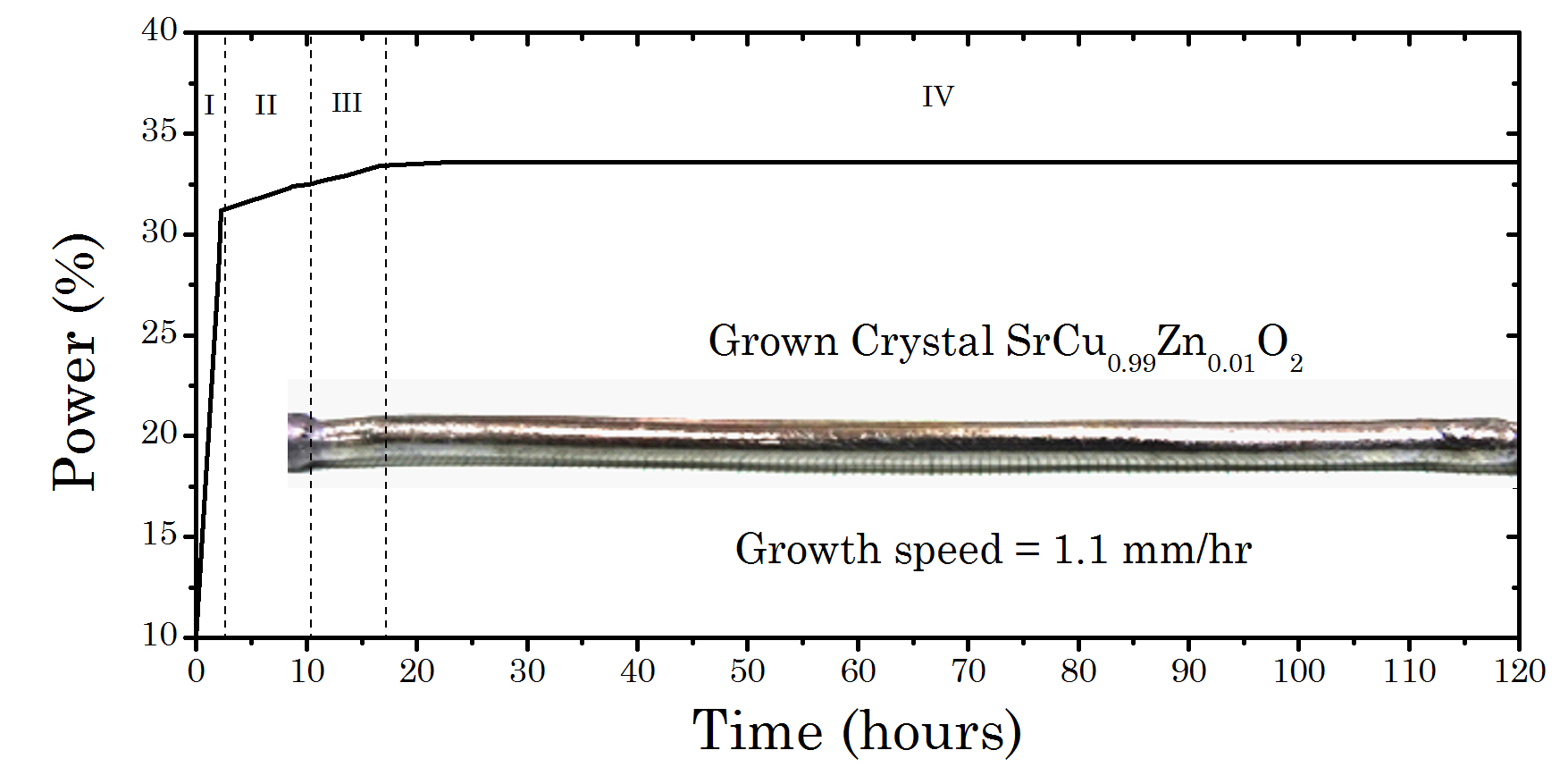}
\caption{Plot of furnace power against time in the TSFZ growth experiment of SrCu$_{0.99}$Zn$_{0.01}$O$_{2}$. Region I: Power increased fast to reach a temperature close to the peritectic temperature. Region II: The solvent disk is partially melted and its temperature is gradually raised to enrich the liquid with Strontium. Region III: Fine tuning of the furnace power to reach the steady state. Region IV: The steady state region. See text for details.}
\label{PvsT}
\end{figure}

In the present work, we overcome this difficulty using a two step process. In the first step, the solvent disk is fused to the lower end of the feed rod. This is achieved by first fixing the feed to the lower shaft, placing the disk on top of it and increasing the furnace power slowly to reach the temperature where the disk, majority of which is kept outside the hot region, starts decomposing  to produce a thin liquid layer at the interface. Upon quickly cooling by decreasing fast the furnace power, the liquid layer freezes bonding the disk temporarily to the feed rod. In the second step, the feed, with the fused disk at its lower end, and the seed are loaded in the furnace in the conventional way. After flushing the growth chamber with O$_{2}$ gas for 12 hours the power is increased in 3-4 hrs to a value close to the power at which the pellet was fused in the previous step. From here onward, a slow automated heating process is started to melt the pellet completely which is then used to form a floating-zone. Because the process temperature cannot be measured directly in an image furnace of the type used in the present work, some uncertainty may remain in the floating-zone composition which may require further fine tuning to reach the steady state. To illustrate this point further, we show in Fig. \ref{PvsT} a plot of furnace power vs. time (halogen lamps of power 500 W each were used) during the growth experiment of SrCu$_{0.99}$Zn$_{0.1}$O$_{2}$. In the graph, the various stages of the TSFZ process are indicated. Region IV corresponds to the steady-state condition, during this phase no power adjustment was required. In the region III, the furnace power is increased from 32.5\% to 33.6\% to fine tune the temperature-composition of the floating-zone. Regions II and I correspond to initial phases, i.e., prior to floating zone formation. During stage I, the power is quickly increased from 0 to 28\% in 2-3 hrs and in stage II (slow-heating regime) the power is increased gradually from 28\% to 32.5\%. With some experience, one can eliminate completely or minimize the stage III  as is evident from the slopes of the heating curves in the two regions. 

Apart from the solvent composition and its melting behaviour, there are other considerations to keep in mind to allow a TSFZ process to proceed uninterruptedly. The density and diameter of the feed rod should be uniform throughout its length, because changes in the zone-volume due to non-uniform melting of the feed rod may eventually lead to a failure of the growth process (e.g., growth SrCu$_{0.0995}$Zn$_{0.005}$O$_{2}$, Table \ref{GrowthProc}). Another complication that can prove detrimental to the stability of the floating zone concerns the off-axis growth, that is, the crystal prefers to grow away from the axis of the furnace as seen in the case of SrCuO$_{2}$-II in Table \ref{GrowthProc}. Typically, slower rotation speed helps in keeping the growing crystal along the furnace axis, but too slow a rotation speed may result in temperature-compositional inhomogeneity within the floating zone. In the present work we typically used a rotation speed of 15 rpm, however, in few cases, where a slight tendency towards off-axis growth was noticed, rotation speeds as low as 6 rpm were used. 

\begin{figure}[hbtp]
\centering
\includegraphics[width=0.6\textwidth]{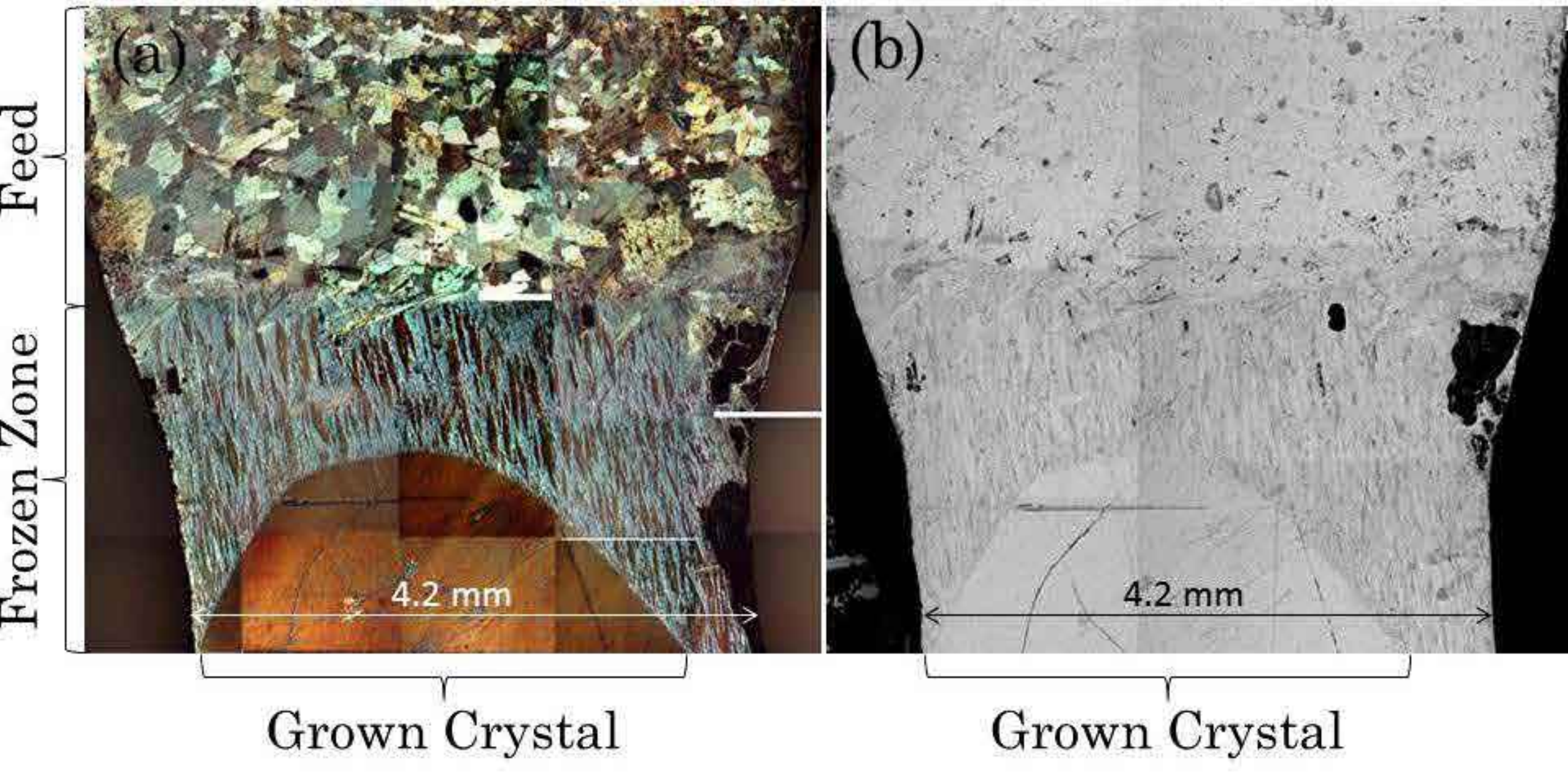}
\caption{(a) Optical image under polarized light of the frozen-in steady state floating zone and (b) the corresponding image obtained using SEM.}
\label{FZ}
\end{figure}

To study the steady state composition of the floating zone during the growth process and the shape of the solid-liquid interface, the floating zone in one of the successful growth experiments on pure SrCuO$_{2}$ was frozen towards the end of the growth by cooling the furnace quickly, which essentially led to quenching of the liquid zone. The longitudinal section of the frozen-in zone was polished and examined using optical microscope and scanning electron microscopy. The results are shown in Fig. \ref{FZ}. The energy dispersive X-ray analysis of the frozen-in zone confirms that its average composition consists of 31.3\% SrO and 68.7\% CuO mixtures, in excellent agreement with the composition of the associated peritectic point \cite{PD2, PD1}. The growth-melt interface under steady state condition exhibits a convex shape as shown in Fig. \ref{FZ}. The interface shape has been argued to be one of the important factors that governs the quality and diameter of a crystal grown using the floating-zone method. A concave interface is considered unfavorable as it tends to enhance the concentration of defects along the core of the growing crystal\cite{Kitamura1}. On the other hand a slightly convex melt-growth interface is considered desirable with the least number of defects reported \cite{Kimura}. In our case, the shape of the interface is convex, however, the convexity appears to be enhanced, which suggests a relatively large temperature gradient at the interface of the growing crystal and the possibility of defects and disorientation in the peripheral regions. Typically, the absorption of infra-red radiation at the surface of the melt results in a small temperature gradient in the radial direction which may be further augmented due to anisotropic thermal conductivity of the growing crystal. Low relative thermal conductivity in the transverse direction of the growing crystal may result in a temperature gradient within the floating zone, with a slightly cooler core region. One can apply faster rotation speeds to counteract the gradient. New growth experiments using higher rotation speed are being planned to understand the shape of the melt-crystal interface and its influence on the quality of the grown crystal. 

Table \ref{GrowthProc} summarizes the results of several growth experiments. The diameters of all the grown crystals range between 4 and 4.5 mm, and their length between 70 and 100 mm. The crystals were found to cleave rather easily with large mirror finished surface whose orientation was later confirmed to be perpendicular to the crystallographic \emph{b}-axis, i.e., the cleavage plane contains the CuO chains [Fig. \ref{crystruc}]. In some cases, where the \emph{ac}-plane of the crystal happens to be oriented along the growth axis, the length of the cleavage section extended over the entire length of the grown crystal boule. The relative ease with which these crystals cleave can be attributed to a weaker inter-planar bonding compared to the in-plane bonding. Within the planes, the magnetic Cu-O-Cu exchange interaction along the chain direction is about $\sim$2000 K, which is far too high compared to the temperature at which their TSFZ growth occurs ($\sim$1370 K). It might, therefore, be possible that their growth behavior is governed by the in-plane magnetic bonding. 

The powder X-ray diffraction data of the sample containing 2.5 \% cobalt showed the presence of small amounts of unidentified phases, which indicates that the actual amount of Co that can substitute for Cu in SrCuO$_{2}$ is less than 2.5 \%. As a result, in the TSFZ experiment of this compound, the floating zone could not be kept stable for more then several hours at a time.    

\subsection{Structural characterization}
Due to their tendency to cleave rather easily along the growth direction, cutting perpendicular sections from the grown crystal boules presented some difficulties. We employed a 30 $\mu$m thickness diamond saw at very slow speeds and without the use of any coolant, to cut perpendicular sections at different places from one of the grown crystal boules (SrCu$_{0.99}$Co$_{0.01}$O$_{2}$ -II). These sections were dry polished using silicon-carbide papers of progressively higher grit sizes with final polishing at the grit size 1200 (approx. 15 microns particle size) and were examined under an optical microscope using polarized light. The results of two different sections  at 44 mm and 76 mm from the beginning of the growth are shown in Fig. 7. The polished section at 44 mm shows segregation of extra phase located along nearly concentric circular arcs. Above 48 mm, these arcs were pushed further out towards the peripheral region of the crystal. The section at 76 mm does not show any signs of segregation. Along the circumference of the crystal, however, smaller regions of inhomogeneity or of differently oriented domains compared to the bulk of the crystal were seen in some of the sections, which is likely the result of large convexity of the melt-growth interface as was discussed above. The X-ray powder diffraction measurements done on several sections indicated the existence of an extra phase of Sr$_{14}$Cu$_{24}$O$_{41}$ in almost all the grown crystals. In the best cases, segregation disappeared after 10-15 mm of growth and in the worst cases it persisted up to about 50 mm of growth length. These variations are related to the composition of the initial floating zone. 

Several freshly cleaved sections from various regions of the grown crystals were rigorously investigated using the SEM. To elucidate the morphology and growth behaviour, some representative secondary electronic images are shown in Fig. \ref{SEM}. The layered morphology of the grown crystals, derived from their anisotropic layered structure, is evident in these pictures  due to the appearance of steps and terraces typically seen on the cleaved sections of layered crystals. The examined sections showed a very homogeneous composition. In particular, none of the examined specimens showed any secondary phases or aggregates, suggesting a uniform distribution of the dopants within the crystal bulk. These findings are further corroborated by magnetization studies presented in the next section. Attempts to determine the actual Co concentration in our crystals using EDX analysis did not prove very successful because their small concentration is almost at the sensitivity limit of the EDX probe. The lattice parameters shown in Table \ref{Latticeparameter} were obtained from powder diffraction patterns using the UNITCELL software. No significant variation of the lattice parameters with Co and Zn doping could be observed, which is expected since the ionic radii differ only slightly in +2 oxidation state and 4-fold coordination \cite{Shannon} for Cu $\simeq$ 58 pm, Co $\simeq$ 57 pm and Zn $\simeq$ 60 pm. The crystallographic orientation of the cleaved surfaces is confirmed using Laue diffraction and X-ray diffraction in the Bragg-Brentano geometry [Fig. \ref{Laue}]. Sharp Laue spots are indicative of high crystal quality and the symmetry of the Laue spots confirms that the cleaved face is the crystallographic \emph{ac} plane. X-ray diffraction from the cleaved faces exhibit sharp lines having indices 0k0, k even. The absence of diffraction lines other than 0k0 also confirm that the crystals cleave along the crystallographic \emph{ac}-plane and that they are free from any secondary phase(s). Since X-ray diffraction is essentially a surface sensitive technique,1.5-2 cm long single crystal pieces cut from the as-grown boule of pure and 1\% Co doped sample were investigated using neutron diffraction to asses the crystalline quality of the bulk. The experiments were performed by fixing the detector at the corresponding Bragg angle and rocking the crystals about their \emph{b} axes. The crystals were fully illuminated by the neutron beam during these experiments. The full-width at half-maximum (FWHM) of the diffraction peak at 001 is found to be around 0.5$^{\circ}$ for both pure and 1 \% Co-doped crystals which again points to a very good crystalline quality (Fig. \ref{rockingcurve}).

\subsection{Magnetization}
The dc susceptibility of a pure SrCuO$_{2}$ crystal measured along the three principal crystallographic directions \textit{a}, \textit{b} and \textit{c} is shown in Fig. \ref{Magneticdata}(a). The magnitudes of the susceptibilities along the three directions and the small temperature independent anisotropy ($\chi_{a} > \chi_{b} \approx \chi_{c}$) are in good agreement with a previous report\cite{Motoyama}. Since the observed anisotropy is temperature independent, arising from the different Van-Vleck paramagnetic contributions along the three directions, the spin part of the susceptibility should be isotropic, as argued by Motoyama et al. \cite{Motoyama}. Therefore, the spins S = $\frac{1}{2}$ of Cu$^{2+}$ in SrCuO$_{2}$ are Heisenberg like. The Zn doped samples exhibit a behavior rather similar to that of the pure compound albeit with a slightly enhanced Curie-tail at low temperature (data not shown), which is attributed to the Zn ions (S = 0) acting as effective chain breaks. In the neighbourhood of these breaks small free paramagnetic moments are understood to lead to the enhancement in the Curie tail\cite{Sirker}. Doping of Co, on the other hand, results in two distinctive changes in the magnetic behavior of SrCuO$_{2}$ (the susceptibility of the 1 \% Co doped crystal is shown as a representative in Fig. \ref{Magneticdata}(b)): (1) while the \textit{a} and \textit{b} axes susceptibilities are comparable to that of the pure compound, along the \textit{c} axis, a large increase is observed at low temperatures, amounting to values about 7 times that of the pure compound along this direction; and (2) a sharp peak occurs in the susceptibility near T = 5 K. The peak temperature is found to scale linearly with the Co concentration. The splitting of zero field cooled (ZFC) and field cooled (FC) susceptibility, signalling minor thermomagnetic irreversibility below the magnetic peak temperature (upper inset, Fig. \ref{Magneticdata}(b)) suggests that Co doped SrCuO$_{2}$ undergoes a spin freezing whose onset temperature scales almost linearly with Co concentration (lower inset, Fig. \ref{Magneticdata}(b)). This is to be contrasted with a recent study where it is shown that comparable amount of Ni doping in SrCuO$_{2}$ results in a pseudo spin-gap opening below T = 50 K \cite{Simutis}. It is interesting to note that drastic changes in the bulk properties are induced by magnetic impurities in the quasi 1D spin chain compound SrCuO$_{2}$. Further studies are being done to understand the observed behaviors.

\section{Conclusion}
In conclusion, we have grown high quality single crystals of pure SrCuO$_{2}$ and its doped variants consisting of Co (0.25\%, 0.5\%, 1\% and 2.5\%) and Zn (0.5\% and 1\%) substitution using the travelling solvent floating-zone method associated with an optical floating zone furnace. We have shown that a stable floating zone that can work for days without interruption can be established right at the beginning of the growth experiment by choosing the composition of the solvent disk to be the same as that of the peritectic point associated with the decomposition of SrCuO$_{2}$ (30 \% Sr : 70 \% Cu) and by paying due considerations to the melting behaviour of the solvent disk itself. The high quality of the grown crystals is ascertain using several complimentary techniques including scanning electron microscopy X-ray diffraction and the rocking curves measured by neutron diffraction on cm-size crystals. The magnetic susceptibility of the pure compound is found to be in excellent agreement with the previous single-crystal report. Preliminary magnetization studies on the Co-doped crystals showed highly anisotropic behaviour with a low-temperature peak in the \emph{c}-axis magnetic susceptibility observed for all Co-doping levels. The temperature at which the peak appears is found to scale linearly with  the Co-content. The upper limit of Cu$^{2+}$ substitution by Co$^{2+}$ seems to be less than 2.5 atomic \%.  

\begin{acknowledgement}
S.S. and C.R. thankfully acknowledge the Department of Science and Technology (DST), India, and the Swiss State Secretariat for Education, Research and Innovation (SERI) for financial support under the Indo-Swiss Joint Research Programme (Grant number: INT/ SWISS/ ISJRP/ PEP/ P-06/ 2012). S.S. and K.K. are thankful to Prof. Sudesh Dhar (TIFR) for many helpful suggestions and help in carrying out some preliminary magnetic measurements at TIFR, Mumbai.
\end{acknowledgement}

\bibliography{Bibliography}

\pagebreak
\begin{table}
\centering
\begin{tabular}{l r r c c p{5cm}}
\hline 
Compounds & \multicolumn{2}{c}{Precursor} & \multicolumn{2}{c}{Heat Treatment} & Final Phase \\ 
   & SrCO$_{3}$ & CuO & Temp($^{\circ}$C) & Hours &   \\ 
\hline 
SrCuO$_{2}$-I & 99.9\% (S) & 99\% (S) & 800-985 & 120 & small extra phase* \\ 
SrCuO$_{2}$-II & 99.9\% (S) & 99.99\% (S) & 900-980 & 145 & single-phase \\ 
SrCu$_{0.995}$Zn$_{0.005}$O$_{2}$ & 99.99\% (A) & 99\% (S) & 825-960 & 400 & single-phase \\
SrCu$_{0.99}$Zn$_{0.01}$O$_{2}$ & 99.9\% (S) & 99.995\% (A) & 800-975 & 150 & single-phase \\ 
SrCu$_{0.9975}$Co$_{0.0025}$O$_{2}$ & 99.9\% (S) & 99.995\% (A) & 800-975 & 190 & small extra phase* \\
SrCu$_{0.995}$Co$_{0.005}$O$_{2}$ & 99.99\% (A) & 99.995\% (A) & 850-975 & 275 & small extra phase* \\
SrCu$_{0.99}$Co$_{0.01}$O$_{2}$-I & 99.9\% (S) & 99\% (S) & 900-990 & 120 & single-phase \\ 
SrCu$_{0.99}$Co$_{0.01}$O$_{2}$-II & 99.9\% (S) & 99.995\% (A) & 850-975 & 210 & single-phase \\
SrCu$_{0.975}$Co$_{0.025}$O$_{2}$ & 99.9\% (S) & 99.995\% (A) & 800-975 & 240 & small unidentified extra phase \\ 
\hline 
\end{tabular}\\
S:Sigma-Aldrich; A:Alfa-Aesar; * Sr$_{14}$Cu$_{24}$O$_{41}$.
\caption{Details of compound synthesis}
\label{Synthesis}
\end{table}
\begin{table}
\centering
\begin{tabular}{l c c l l p{5cm}}
\hline
Compounds & \multicolumn{2}{c}{Growth rate(mm/h)} & Disk & Growth & Remarks \\
 & v$_{f}$* & v$_{g}$** & SrO:CuO & length(mm) &  \\ 
\hline 
SrCuO$_{2}$-I & 0.92 & 1.00 & 3 : 7 & 79.5 & Stable Float Zone (FZ) \\ 
SrCuO$_{2}$-II & 0.80 & 1.10 & 3 : 7 & 82.2 & FZ broke due to off-axis growth \\ 
SrCu$_{0.995}$Zn$_{0.005}$O$_{2}$ & 0.80 & 1.00 & 3 : 7 & 71.4 & Unstable FZ due to non-uniform diameter of the feed\\
SrCu$_{0.99}$Zn$_{0.01}$O$_{2}$ & 0.85 & 1.00 & 3 : 7 & 81.8 & Stable FZ \\
SrCu$_{0.9975}$Co$_{0.0025}$O$_{2}$ & 0.92 & 1.00 & 3 : 7 & 115 & Stable FZ \\
SrCu$_{0.995}$Co$_{0.005}$O$_{2}$ & 0.92 & 1.00 & 3 : 7 & 80.1 & Stable FZ \\
SrCu$_{0.99}$Co$_{0.01}$O$_{2}$-I & 0.85 & 1.05 & 3 : 7 & 81.3 & Stable FZ \\
SrCu$_{0.99}$Co$_{0.01}$O$_{2}$-II & 0.80 & 1.05 & 3 : 7 & 125 & Stable FZ \\
SrCu$_{0.975}$Co$_{0.025}$O$_{2}$ & 0.80 & 1.00 & 14 : 24 & 80.0 & Unstable FZ due to extra phase \\
\hline 
\end{tabular}\\
* Feeding speed; ** Growth speed.
\caption{Details of growth process}
\label{GrowthProc}
\end{table}
\begin{table}
\centering
\begin{tabular}{l c c c c}
\hline 
Sample & a & b & c & Cell volume \\ 
\hline 
Standard\cite{Matsushita} & 3.577 & 16.342 & 3.918 & 229.028 \\ 
Pure & 3.574 & 16.332 & 3.911 & 228.310 \\ 
Zn0.5\% & 3.572 & 16.328 & 3.910 & 228.063 \\
Zn1\% & 3.572 & 16.333 & 3.910 & 228.118 \\ 
Co0.25\% & 3.575 & 16.337 & 3.911 & 228.451 \\
Co0.5\% & 3.571 & 16.322 & 3.909 & 227.825 \\ 
Co1\% & 3.574 & 16.331 & 3.912 & 228.277 \\
Co2.5\% & 3.574 & 16.335 & 3.910 & 228.295\\ 
\hline 
\end{tabular}
\caption{Lattice parameter for the grown crystals}
\label{Latticeparameter}
\end{table}

\begin{figure}[hbtp]
\centering
\includegraphics[width=0.8\textwidth]{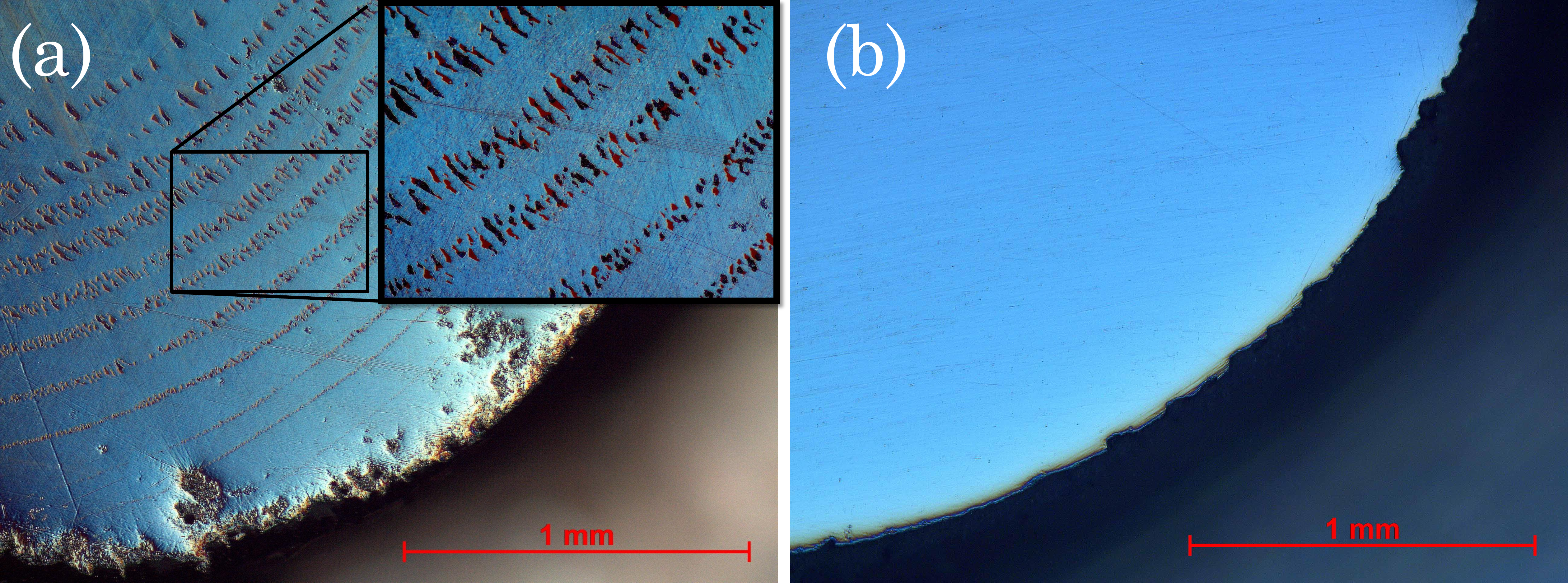}
\caption{Representative optical images under polarised light of the polished cross-sections at different lengths of the grown crystal (SrCu$_{0.99}$Co$_{0.01}$O$_{2}$-III) (a) at 44 mm and (b) at 78 mm from the beginning of the growth.}
\label{PhaseSeg}
\end{figure}

\begin{figure}[hbtp]
\centering
\includegraphics[width=0.4\textwidth]{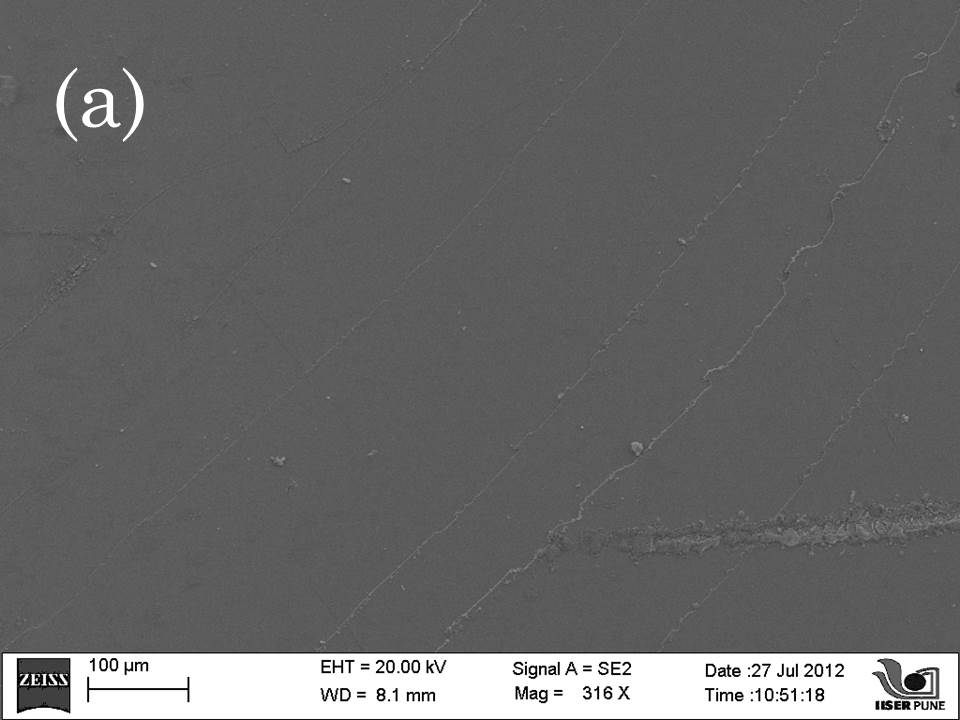}
\includegraphics[width=0.4\textwidth]{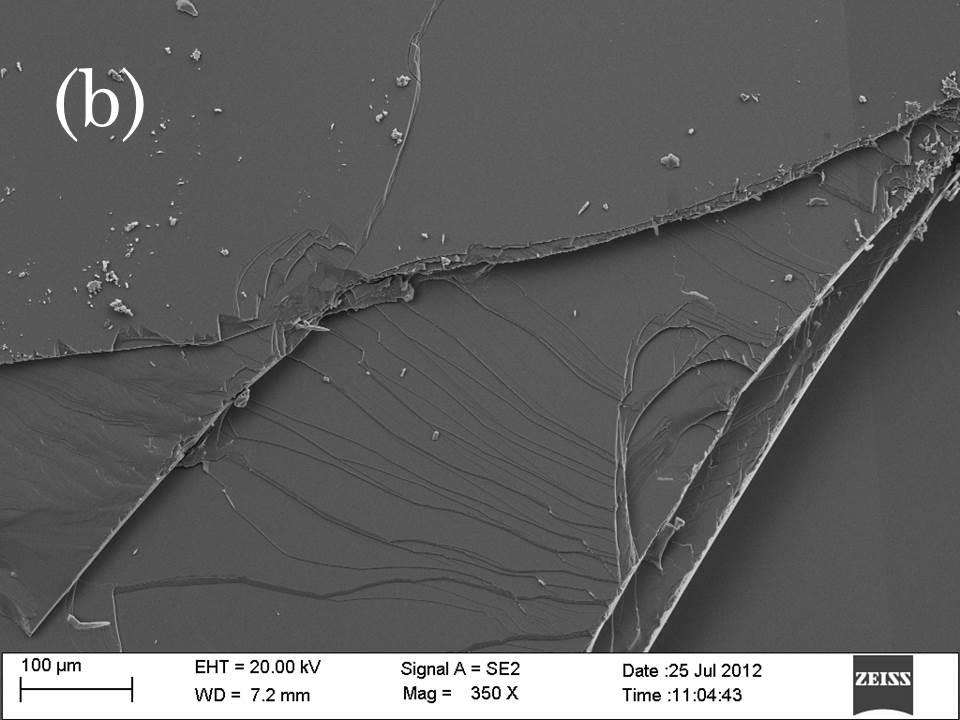}
\caption{The SEM images of (a)Zn1\% (b)Co1\%.}
\label{SEM}
\end{figure}

\begin{figure}[hbtp]
\centering
\includegraphics[width=0.8\textwidth]{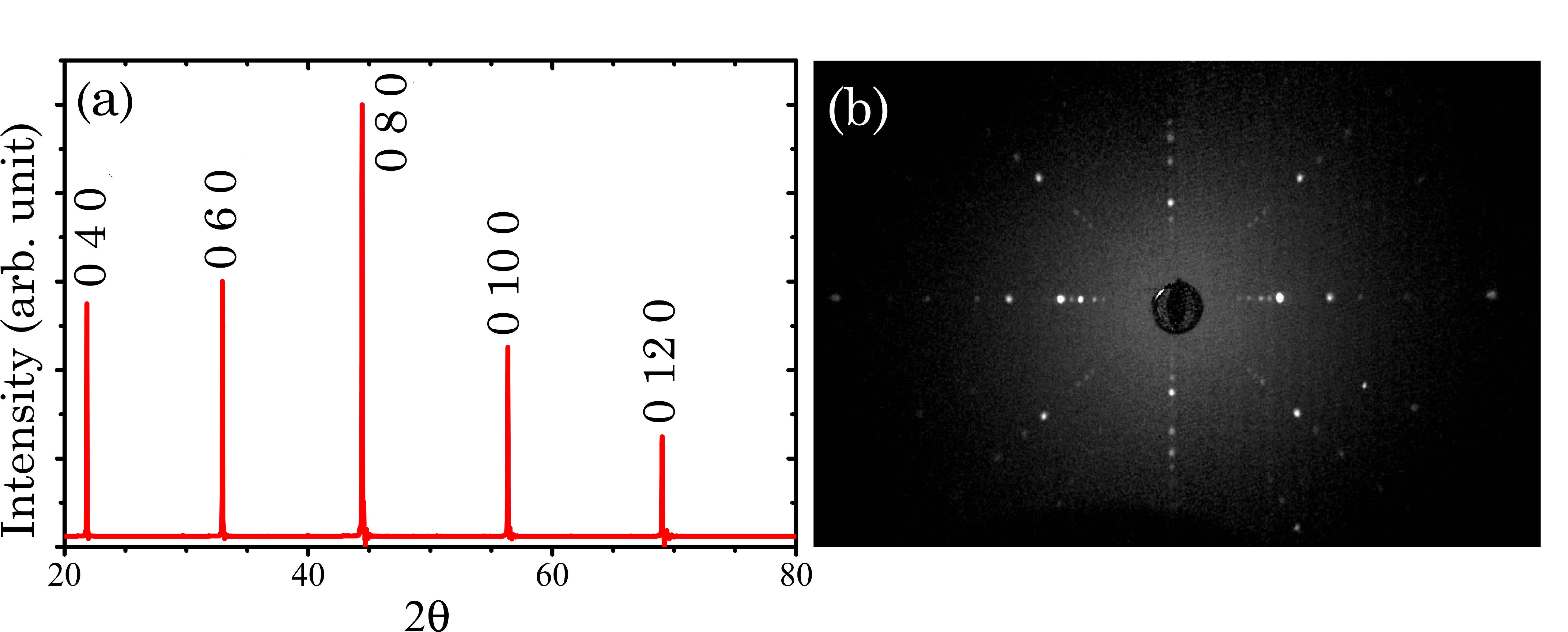}
\caption{(a) The XRD pattern from a cleaved surface of SrCuO$_{2}$ using Bragg-Brentano geometry, and corresponding Laue diffraction pattern (b).}
\label{Laue}
\end{figure}

\begin{figure}[hbtp]
\centering
\includegraphics[width=0.4\textwidth]{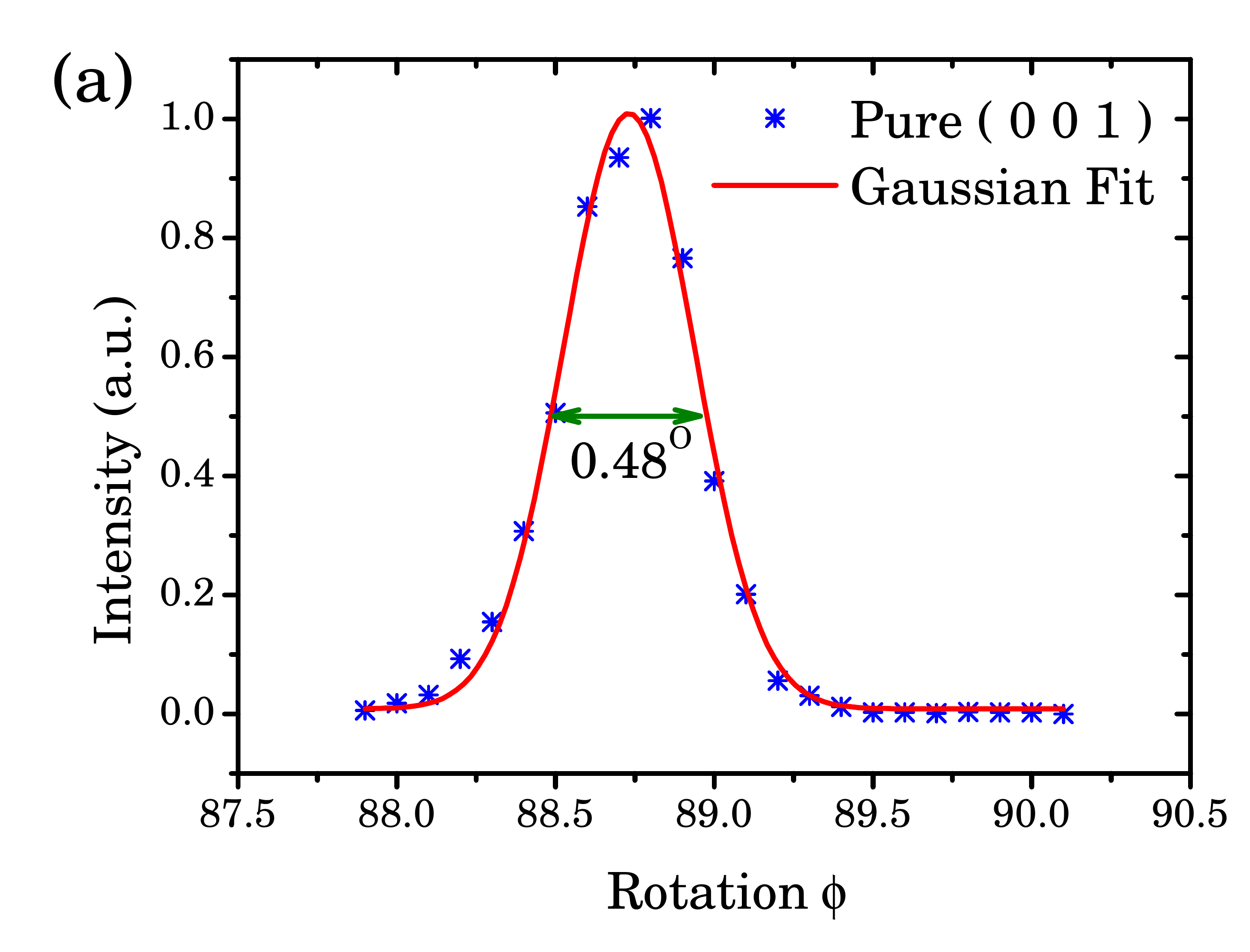} \includegraphics[width=0.4\textwidth]{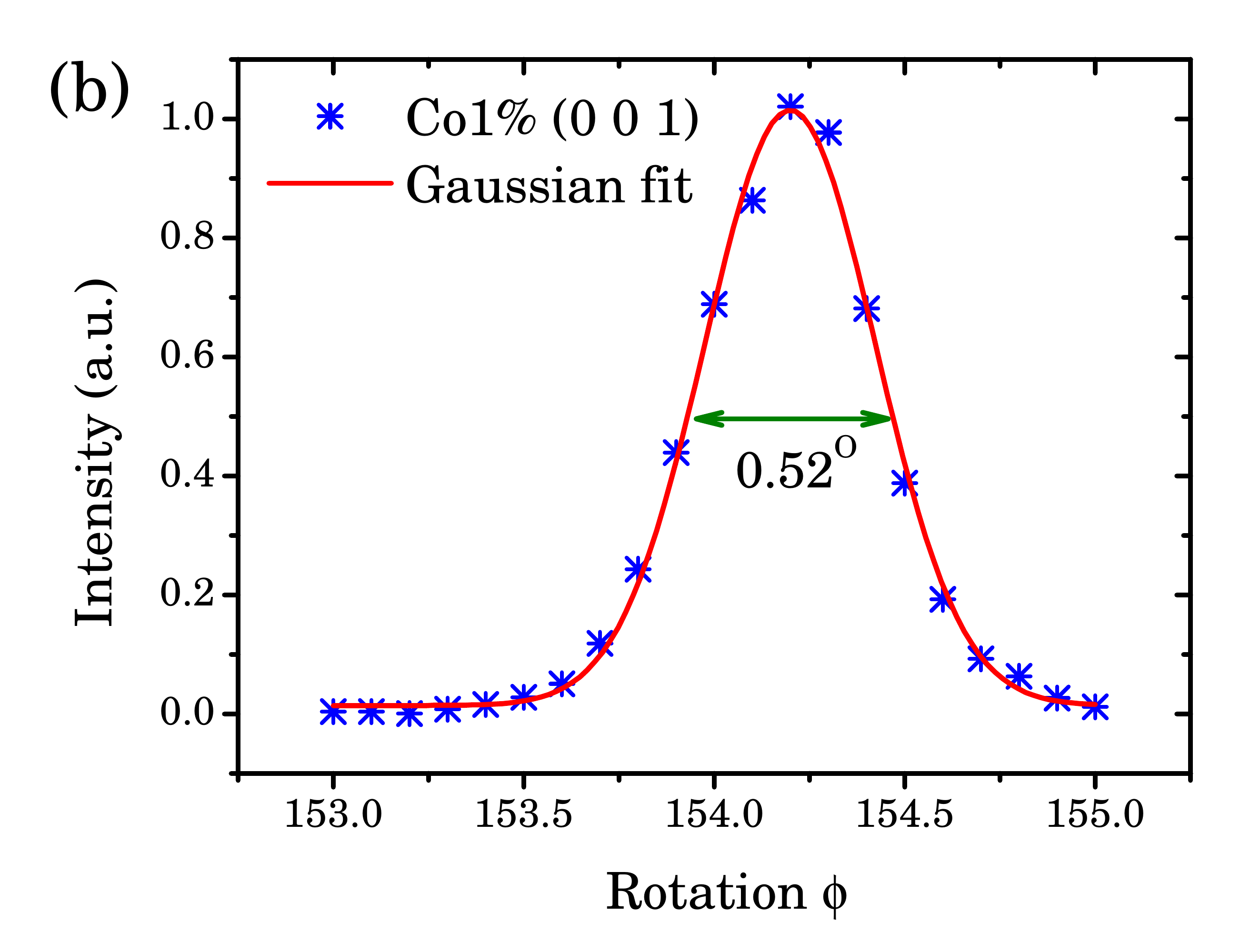}
\caption{Rocking curve of nuclear Bragg reflections measured by neutron diffraction (a) pure (b) Co1\% doped single crystals (see text for details). The solid lines are Gaussian fits to the data.}
\label{rockingcurve}
\end{figure}

\begin{figure}[hbtp]
\centering
\includegraphics[width=0.8\textwidth]{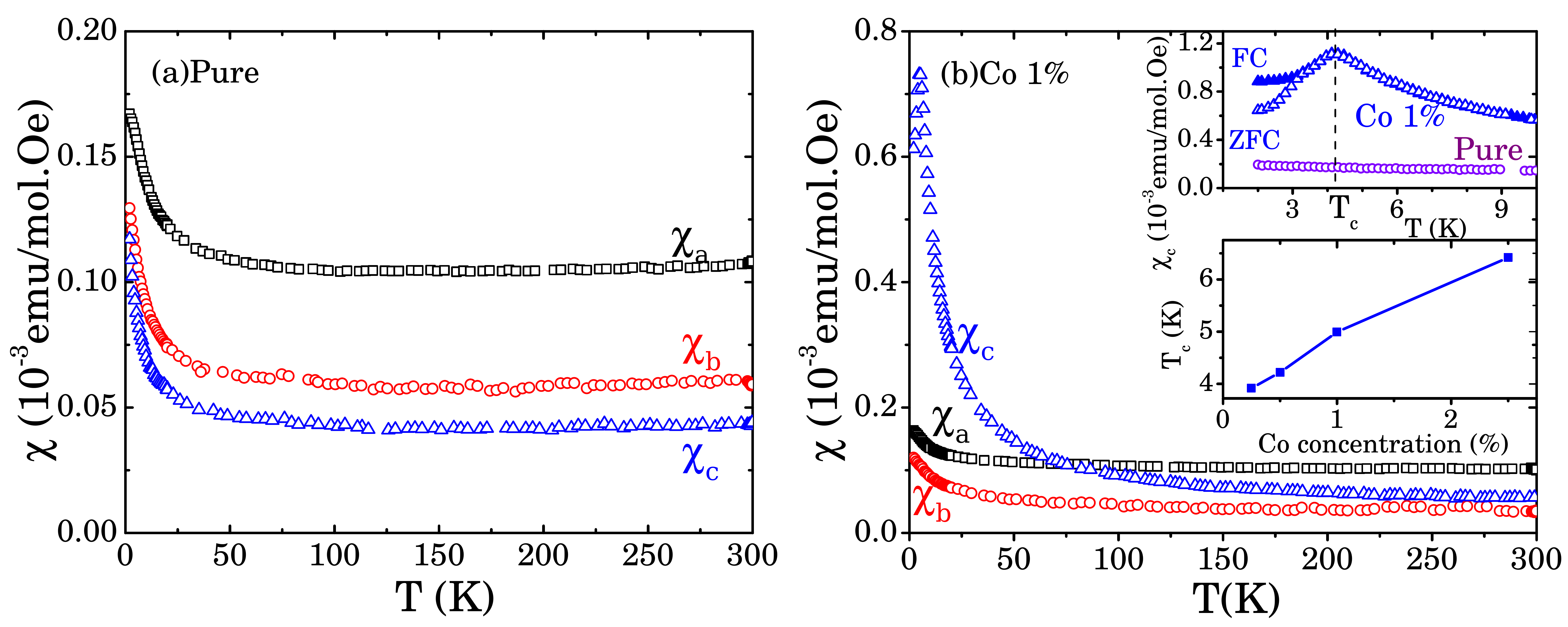}
\caption{Temperature dependent magnetic susceptibility ($\chi$) of (a) pure and (b)1\% Co-doped crystals along the crystallographic \emph{a}, \emph{b} and \emph{c} axis (H = 10kOe). The upper inset shows the low-temperature susceptibility ($\chi_{c}$ in H = 500 Oe) of the pure and Co1\% doped crystals. The lower inset shows the variation of the peak temperature in $\chi_{c}$ (H = 10kOe) for various Co-concentrations.}
\label{Magneticdata}
\end{figure}

\end{document}